# Toward Cybersecurity Testing and Monitoring of IoT Ecosystems

Steve Taylor, University of Southampton, sjt1@soton.ac.uk, https://orcid.org/0000-0002-9937-1762

Panos Melas, University of Southampton, pmelas@soton.ac.uk, https://orcid.org/0000-0002-2825-9124

Martin Gile Jaatun, SINTEF, Martin.G.Jaatun@sintef.no, ORCID ID 0000-0001-7127-6694

Aida Omerovic, SINTEF, aida.omerovic@sintef.no

Robert Seidl, Nokia Bell Labs, robert.seidl@nokia-bell-labs.com, ORCID ID 0000-0001-8518-8433

Norbert Goetze, Nokia Bell Labs, norbert.goetze@nokia-bell-labs.com, ORCID ID 0009-0009-6277-6916

Jens Kuhr, Nokia Bell Labs, jens.kuhr@nokia.com, ORCID iD 0009-0003-6818-9380

Dmytro Prosvirin, Antonov Airlines, dmytro.prosvirin@antonov-airlines.aero

Manuel Leone, Telecom Italia, manuel.leone@telecomitalia.it, ORCID ID 0009-0000-3707-2882

Paolo De Lutiis, Telecom Italia, paolo.delutiis@telecomitalia.it, ORCID 0009-0001-8842-8348

Andrey Kuznetsov, WRCVE, wrcve3000@gmail.com

Anatoliy Gritskevich, WRCVE, anatolii.hrits@gmail.com

George N. Triantafyllou, ATC, georgestri@yahoo.com, ORCID ID 0000-0001-5322-206X

Antonis Mpantis, ATC, a.mpantis@atc.gr, ORCID ID  0009-0001-1304-530X

Oscar Garcia Perales, i4RI, oscar.garcia@i4ri.com, ORCID ID 0009-0003-0781-6141

Bernd-Ludwig Wenning, MTU, Email: berndludwig.wenning@mtu.ie , ORCID ID  0000-0002-2034-253X

Sayon Duttagupta, COSIC, KU Leuven, Sayon.Duttagupta@esat.kuleuven.be, ORCID ID: 0000-0002-3495-4641

## Abstract

We describe a framework and tool specification that represents a step towards cybersecurity testing and monitoring of IoT ecosystems. We begin with challenges from a previous paper and discuss an integrated approach and tools to enable testing and monitoring to address these challenges. We also describe exemplary use cases of IoT ecosystems and propose approaches to address the challenges using the framework and tools. The current status of this work is that the specification and conceptualisation is complete, use cases are understood with clear challenges and implementation / extension of the tools and framework is underway with tools at different stages of development. Several key observations have been made throughout this work, as follows. 1) Tools may be used in multiple different combinations, and ad-hoc use is also encouraged, where one tool may provide clues and other tools executed to undertake further investigations based on initial results.  2) Automated execution of tool chains is supported by workflows. 3) support for immutable storage of audit records of tests and results is an important requirement. 4) Indicators (observations or measurements representing information of relevance for assessment of cyber security) are a key mechanism for intercommunication between one tool and another, or with the operator. 5) Mapping this work to established security development lifecycles is a useful means of determining applicability and utility of the tools and framework. 6) There is a key interplay between devices and systems. 7) Anomaly detection in multiple forms is a key means of runtime monitoring. 8) Considerable investigation is needed related to the specifics of each device / system as an item of further work.

# Introduction

This paper follows on from Taylor et al. (2024), which described challenges for the cybersecurity testing and monitoring of IoT ecosystems and proposed a high-level approach to address these challenges. This paper provides more development details of a framework, tools and methodologies currently in development aiming to address these challenges. As such, it describes the status in the Horizon Europe TELEMETRY project after one year of its three-year duration. The overall architecture of the framework is described, along with the current status of testing and monitoring tools within it. Approaches for addressing the requirements of the use cases are proposed and will be evaluated in three use cases, which are described.

The paper is structured as follows. Next is a background & challenges section, mostly using material from Taylor et al. (2024), as it describes the problem statement this paper aims to address. Following this we describe the methodology and architecture of the testing framework in development, along with details of each tool and component within it. We then briefly describe the three use cases within the TELEMETRY project and show how the tools and the framework may be used in each scenario. Following this is a discussion section that summarises the approaches to addressing the presented challenges, along with key observations made as a result of this work. Finally, brief conclusions summarise the current status and provide pointers for further work.

# Background & Challenges

The societal and economic benefits from the rapid advance of the digital economies are at the core of the European Digital Agenda. This is bolstered by the Next Generation Internet (NGI) initiative with an emphasis on creating a more resilient, trustworthy and sustainable Internet for our digital future. As the Internet of Things (IoT) brings connectivity and networked intelligence to the physical things around us, highly distributed and complex infrastructures are emerging ultimately forming IoT ecosystems, which can be defined as: *systems of interconnected IoT devices with hardware, software, services and backbone network communication infrastructure to support the required system functionality*.

While these ecosystems bring many benefits and efficiencies, their inherent complexity, heterogeneity and dynamicity and distributed nature create challenges for the management of security, testing, validation, reliability and assurance at scale. The characteristics of IoT ecosystems encapsulate some key challenges for cybersecurity testing & assurance of hardware, software and service components such as: i) IoT devices are often "black boxes" to deployers and users, meaning that their structure and inner components are not accessible for testing; ii) IoT devices are hard to update due to the specific nature of their firmware and that the devices may be manifold and geographically distributed and iii) vulnerabilities in one component may allow threats and risks to propagate to other components in a given system.

To address these challenges, there is a need for tools, techniques and holistic methodologies for cybersecurity testing and vulnerability detection at both component level and also in the systems the components are integrated into (Figure 1). This will enable continuous assessment of IoT components and ecosystems over their whole lifecycle, supporting the propagation of assurance for component developers, system integrators and operators, who act on behalf of ecosystem's eventual users.

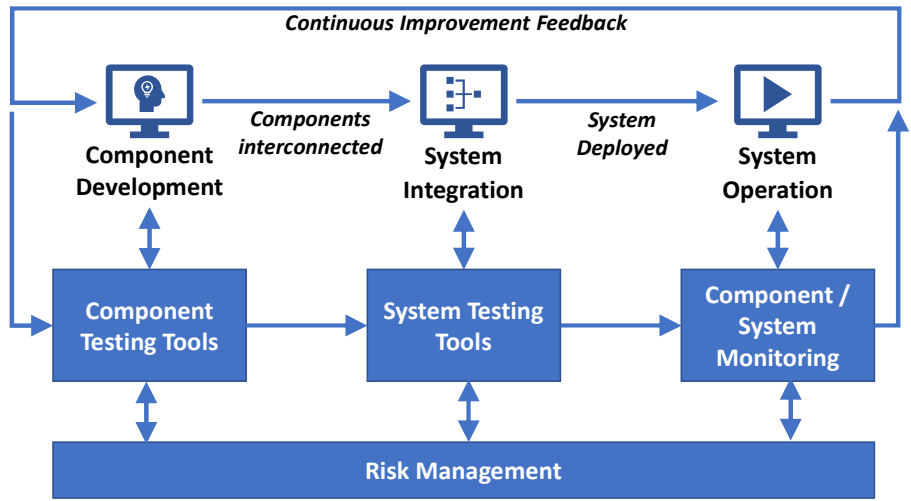

**Figure 1:** *IoT Ecosystem Lifecycle & Testing Tools (reproduced from Taylor et al, 2024).*

We aim to create reliable tools, techniques, and holistic methodologies for cybersecurity testing and vulnerability detection at both component and system level. TELEMETRY will showcase these advancements through three real-world use cases in aviation, smart manufacturing, and telecommunications. The shift towards license-free open-source code, cloud services, and distributed open networks highlights the importance of TELEMETRY's tools and techniques for managing complex systems. The project emphasizes the consortium's commitment to enhancing cybersecurity across diverse industrial sectors.

The key observations regarding the design of the framework from Taylor et al. (2024) are summarised as follows and represent targets for the work reported here.

1) There is a need to consider the **full lifecycle of IoT components** – at their design time, their integration into systems, and operation of those systems.

2) **Threats and risks can propagate when components are connected together in systems** - vulnerabilities in one component can affect other components in a system.

3) **IoT devices present limitations to current testing and management** due to geographical distribution, opacity and limited processing power.

4) **Risk assessment fulfils an important requirement** because it enables assessment of what elements are important to the system's stakeholders, how these elements may be compromised, and how the compromises may be controlled.

5) Feedback from operational monitoring of IoT devices can inform firmware updates / patches to the devices but there is a significant challenge in **rolling out these patches to multiple low-power devices geographically distributed**.

# Methodology & Architecture

## Device Under Test vs System Under Test

A key element of TELEMETRY's concept is the relationship between Devices and Systems, as they are the subjects of testing and monitoring of TELEMETRY. For the scope of TELEMETRY, when we consider a Device, it may be a software component, or an IoT device with hardware and firmware. Also for the scope of TELEMETRY, the term "System" may be defined as *"a set of connected things or devices that operate together"*[1]. This implies that Systems are composed of Devices, which is certainly true, in that a Device may be interconnect and interact with other components in a wider System, for example an IP Camera working in a manufacturing environment interacting with a manufacturing robot in a smart factory System. However, given that IoT and software components are themselves composed of sub-components, e.g. hardware, firmware and software, which is highly likely to be built from third party libraries, it is reasonable to consider Devices as Systems in their own right, and the terms "Device" and "System" are applicable from the perspective of the context of concern and visibility. We therefore use the terms "Device Under Test" (DUT) and "System Under Test" (SUT) to describe the subjects of concern. In summary:

- A system is an interconnected set of components
- A device can be a component within a larger system
- A device can be a system in its own right

---

[1] https://dictionary.cambridge.org/dictionary/english/system

In some cases, a Device is a black box (e.g. when an IP Camera is a device bought from a supplier and deployed in a smart factory system – as in TELEMETRY use case 2 – see later). We are often unable to understand, monitor or control the inner workings of the device and in many cases we have to trust the manufacturer. Devices such as IP cameras are often cheap commodity goods with no guarantees of freedom from vulnerabilities or of any software updates should vulnerabilities be discovered. This has given rise to one of the challenges from Taylor (2024) – that the IoT devices are deployed within systems we are concerned about, but we have often little control over their vulnerabilities, and they may compromise other devices in the system. The notion of Software Bills of Materials (SBOMs) is certainly helpful here, as it helps a system deployer to understand the software composition of a device within their system, but the critical challenge is that SBOMs are often not provided and are difficult to create by the user for some devices. We may have control over other aspects of the system, for example we can monitor the network traffic from the IP camera and look for anomalous behaviour such as data transmission to unexpected destinations and block them if detected.

In other cases, if we are building the IP Camera, we are concerned with its composition of its sub-components, e.g. third party libraries, bespoke code and hardware and their interactions so as to perform the function of the camera. Here, we are considering the IP Camera as a System in its own right. If we have this degree of visibility, especially if we are the creator of the device, we will have made decisions about, and therefore understand, the supply chain of software and hardware that the device is composed of.

## Support for Security Development Lifecycle

The Microsoft Security Development Lifecycle (SDL) exemplifies a modern approach to security by design and in operation. The white paper on evolving the SDL (Ornstein and Rice, 2024) describes the lifecycle of the SDL. The concept of TELEMETERY (shown in Figure 1), plus the function of the TELEMETRY tools and infrastructure directly supports the SDL, as described via the relationship between the TELEMETRY approach and the SDL, which is shown in Figure 2.

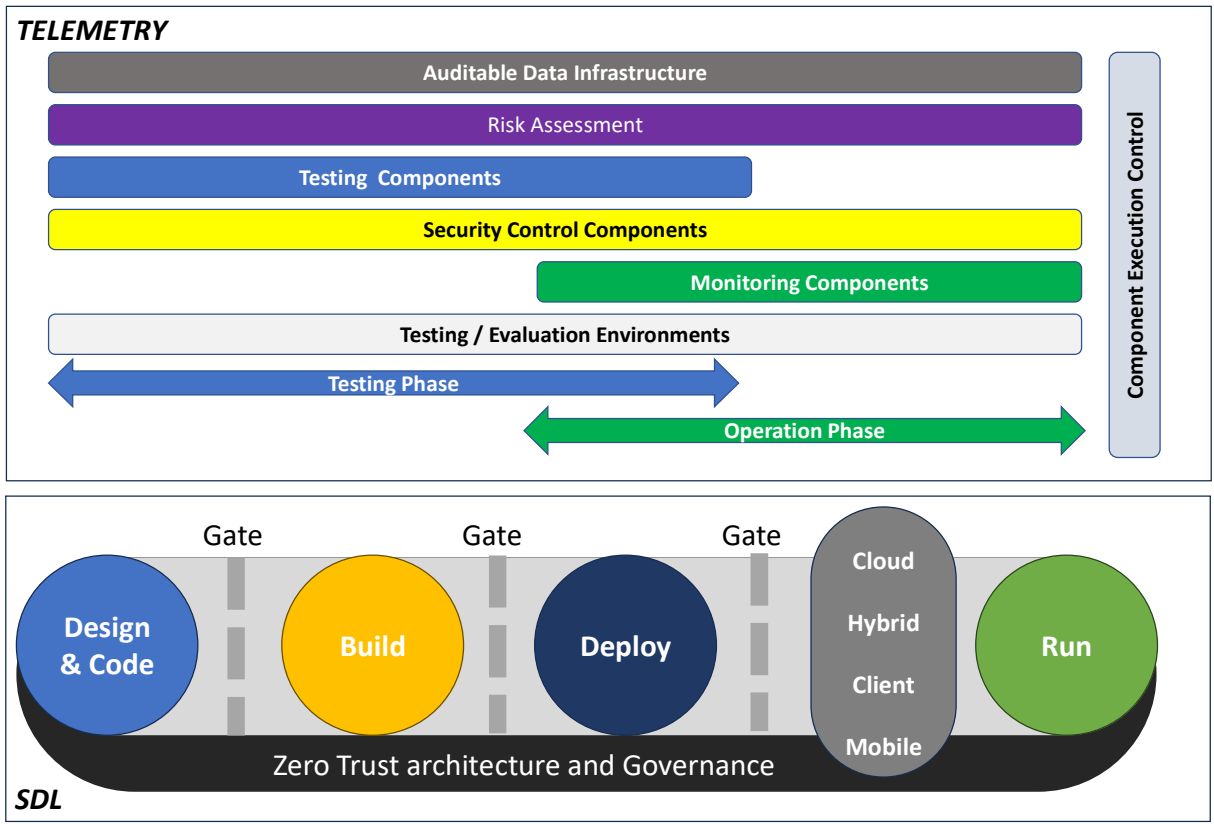

Figure 2: TELEMETRY & Microsoft SDL – SDL reproduced from Ornstein and Rice (2024)

The early phases of the SDL are where the device is engineered (Designed and Built). Depending on the phase, the software may not yet exist (requirements, design), be incomplete (security testing), or completely (or as near as) finished (penetration testing). This can include "conformance testing" of third-party hardware before inclusion in a system. The latter phases of the SDL involve Deployment of devices into systems and operation (Running) of these systems.

## TELEMETRY Architecture

TELEMETRY supports the SDL via different tools & infrastructure that operate at different lifecycle stages of the DUT / SUT, fall into the following classes. TELEMETRY provides tools in each class, but additional third party tools may be integrated into the framework. The TELEMETRY Conceptual Architecture is shown in Figure 3.

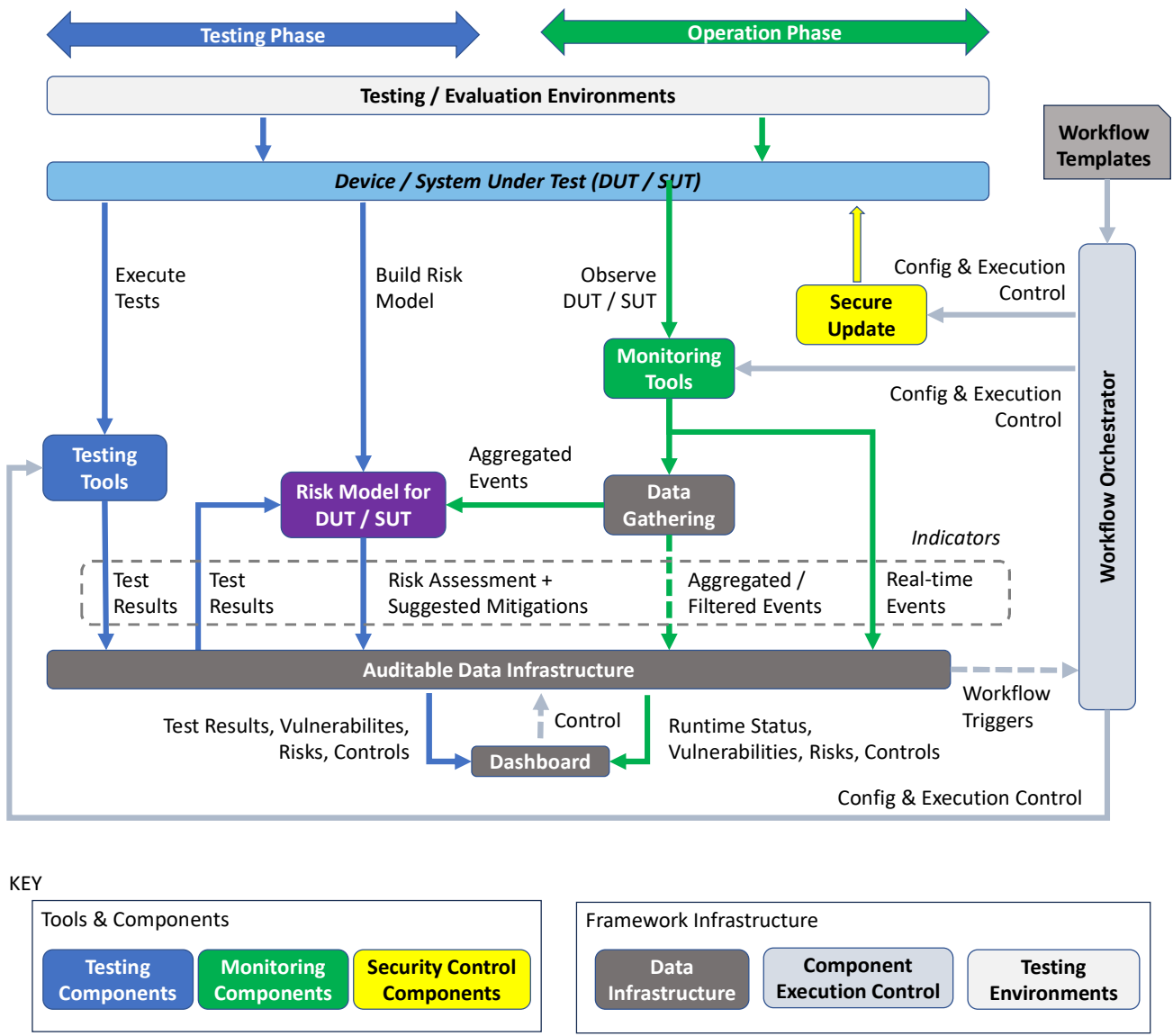

Figure 3: TELEMETRY Conceptual Architecture

**Testing Tools** (blue) are deliberately executed to test some characteristic of the DUT / SUT and with the expectation of specific outputs. These tools are typically used within the Design & Build phases for a device and its Deployment within a wider system, although testing may be undertaken within an operational system in the Run phase also. Examples of TELEMETRY testing tools include Network Fuzzing, SBOM generation and Access Control testing.

**Monitoring, Analysis and Detection Tools** (green) observe the DUT / SUT as it is operating and raise events if specified conditions or anomalous conditions occur that enable us to detect bugs and flaws that we can feed back to the development phase for rectification - "feedback from the field" (McGraw, 2004). These tools are predominantly used with the Run phase of a system with devices, and may generate events that can be communicated to other TELEMETRY tools, such as Risk Assessment or inform additional testing. Such tools may include post-deployment penetration testing tools, intrusion detection tools, and various monitoring tools.

**Security Control Components** (yellow) may be applied to the DUT / SUT to manage risks identified. These components may be used at any stage of the SDL from Desing to Run, depending on the type of control. There are many examples of controls, from best practices to technical measures and TELEMETRY's focus in this area is focused on a technical mechanism for distribution of updates, so is primarily focused within the Run phase of the SDL.

***Risk Assessment*** (purple) evaluates the impact and likelihood of compromises detected by the testing or monitoring & detection tools on the DUT or the SUT, along with recommendations of controls if the resulting risk level is unacceptably high. This can occur at any phase of the SDL and may consider device as a system, where, for example vulnerabilities in third party libraries or hardware affect the output of bespoke code. This is more likely in the earlier stages of the SDL, i.e. Design or Build. Risk assessment may also consider systems of devices at deployment time, i.e. when the system of devices is constructed, in which case, it functions is deliberately executed as a testing too. Alternatively risk assessment may operate at Runtime where it may be informed by the output of monitoring or detection tools, to reactively compute any change in the risk of the operating device or system based on detected events.

***Testing / Evaluation Environments*** – (light grey) are dedicated testbeds, cyber ranges, emulation environments that enable testing under controlled but conditions representative of real deployment conditions. These are typically used at the earlier phases of the SDL (e.g. Design & Build) because they represent testing and evaluation environments as opposed to the real deployment environment, but may also be utilised in later phases (Deploy and Run), especially if they represent Digital Twins of the real environment to enable analysis of dynamic events that occur at operation time.

***Auditable Data infrastructure*** (dark grey) provides means to gather events, to aggregate those events, for tools and other components to securely and auditably exchange information and to interact with testing users. This is a key infrastructural element that operates at all phases of the SDL, as it supports both testing and monitoring / detection.

***Component Execution Control*** (light blue-grey) is infrastructure that enables tools and other components to be configured and executed in different operational sequences depending on the needs of the situation at hand. This will be applicable at all phases of SDL.

***Indicators*** represent information of relevance for assessment of cyber security risk, such as observable vulnerabilities, incidents and threats. Based on initial investigations, indicators have the following properties. Indicators are all types of signal that may trigger an action or be used as evidence in a decision. Indicators can come from each tool and may be used as input for another. Indicators may be aggregated to provide empirical evidence, which is a composition of several pre-defined indicators, for example, one event may not generate an important alarm, but the conjunction and correlation of different events cause an important alarm to be raised. Indicators are likely subject to change over time, due to a change in the SUT / DUT. The changed values need to be captured in timely way, in order to provide up to date information, and the trend over time may in itself be an indicator. Moreover, indicator values may need to be measured on demand in order to confirm an unchanged or assumed value (or state), and thus reduce uncertainty of a risk model.

# Framework Tools & Components

The tools & components of the TELEMETRY approach are described in this section. We use a standardised format for each tool that describes its purpose, features & benefits, inputs & outputs, plus a brief discussion of current status and lessons learned through development of the tool.

## Testing Tools

### Network Fuzzer

The network fuzzer (following e.g. Miller et al., 1990) facilitates security testing of network interfaces, by assisting with the detection of unknown vulnerabilities. The tool achieves this by sending a large amount of specifically crafted requests to the interface under test, and observing whether it responds or behaves in an unexpected way. Such unexpected behaviour can indicate the presence of a vulnerability, which an analyst in turn can investigate further.

**Tool Name:** Network fuzzer. **Purpose:** The network fuzzer facilitates security testing of network interfaces, by assisting with the detection of unknown vulnerabilities. **Features and Benefits:** The main benefit of the tool is that it allows for an automatic detection of unexpected behaviour. By using this as a starting point, security analysts can greatly improve the efficiency of searching for new vulnerabilities. **Uniqueness:** The tool builds on the open source boofuzz[2], which provides a well-documented framework for network fuzzing. Key extensions involve enabling the tool to generate test cases based on network captures from the interfaces of interest, reducing the need for manual configuration and set up, and Implementing a standardized way of delivering/presenting test results. **Input:** A packet capture (.pcap) file[3] from the device or interface to be tested. This file shows the packets and protocols involved with communication to the interfaces which we would like to test. **Output:** The output of the tool should be a set of tuples, containing a description of the observed behaviour and an associated packet capture file with the packets required for reproducing the reported behaviour. **Current Status:** We have employed the network fuzzer in the Telco use case (UC3), both on the generic open source OpenWRT, and the specific device used by Telecom Italia. **Lessons Learned:** Use of the tool still requires significant manual configuration effort on part of the tester, but in the right hands it produces very useful results. It was also interesting to note that among the differences between OpenWRT and the commercial Residential Gateway, the presence of an additional service was also the source of the principal vulnerability discovered, which was previously unreported.

## SBOM Generator

A Software Bill of Materials (SBOM) is a structured overview of all external libraries or software components used to build a software program/system (see Jaatun et al, 2023). There are currently three major SBOM formats: SPDX[4] from the Linux Foundation; CycloneDX[5] from OWASP; and SWID as defined in ISO/IEC 19770-2[6]. The general idea is that software developers should create a distinct SBOM for every version of their software that they publish, enabling customers and users to quickly determine whether, e.g., a given vulnerability applies to the version they are using.

Even though provision of SBOMs has been mandated by the US Government[7] and mentioned in EU legislation[8], they are still far from ubiquitous and many software lacks SBOMs. We have thus explored to what extent we can create an "aftermarket" SBOM based only on analysis of a binary executable or firmware. We also map the discovered versions to the National Vulnerability Database[9] in order to flag any relevant vulnerabilities. This mapping can be performed periodically to catch any newly discovered vulnerabilities.

The SBOM generator gives an overview of the software components and libraries included in a software product. This will in turn allow the tool to list known vulnerabilities present in the software product and the vulnerabilities' severity.

**Tool Name:** SBOM generator. **Purpose:** Provides an overview of the software components and libraries included in a software product. **Features and Benefits:** The purpose of constructing an SBOM is to have an overview of the software components and libraries which are included in a given software product. Having such an overview is a prerequisite for determining which vulnerabilities affect your system, which in turn is essential when performing risk assessments or vulnerability management. **Uniqueness:** While an SBOM is normally generated during the build process of a software product, this tool seeks to perform the same activity after the software has been released, only using the software package or binary file. **Input:** The input to the tool is a binary file or a software package, in the same format as it is delivered by the software development company. **Output:** The output of the tool is an SBOM file, although it is likely not complete. Since the tool effectively performs reverse engineering, 100% coverage should not be expected. **Current Status:** SBOM generation is currently still a predominantly manual process using the package manager available on the firmware of the SUT. The mapping of library versions to CVEs in the NVD is however an automatic process. **Lessons Learned:** The degree of success when applying the SBOM tool depends in a large extent on the properties of the SUT. In the cases where a full SBOM is provided by the manufacturer, the automatic CVE mapping feature will have a close to complete coverage.

---

[2] https://github.com/jtpereyda/boofuzz
[3] e.g. https://www.comparitech.com/net-admin/pcap-guide/
[4] https://spdx.dev/
[5] https://cyclonedx.org/
[6] https://www.iso.org/standard/65666.html
[7] e.g. https://www.infosecurity-magazine.com/news/us-government-proposes-sbom-rules/
[8] e.g https://medium.com/@interlynkblog/eu-cra-and-sbom-5100c55752fa
[9] https://nvd.nist.gov/

## IT infrastructure Access Control Risk Evaluation Using Fuzzy Logic

Complex IT infrastructures may require a single comprehensive access control solution. This is especially relevant for networks that were scaled unevenly, without a pre-designed architecture, with a change in the main responsible in the process, as well as a number of software and administrative solutions that do not make up an optimal and coordinated system. A methodology is proposed, according to which Subjects (users) and Objects (services) are evaluated according to significant factors and with the help of a mathematical model based on fuzzy logic, the risk of providing access is assessed. Bakurova et al. (2024) and Lytvyn et al. (2024) provide full details of the approach, which is intended to help make more informed situational or system decisions for access management.

In this approach, values of indicators and influence factors input are transformed into a set of "if-then" rules as output, where each rule establishes a correspondence between the factors and the level of risk. A set of such rules forms the knowledge base of the system. The values of the factors are processed by these rules, and at the output the model provides an integrated risk assessment. A set of significant factors is suggested in Bakurova et al. (2024) and Lytvyn et al. (2024) but can be modified by the administrator of the SUT. Each factor has 3 to 5 levels of significance, which are determined based on standards (such as CIS Benchmark, CVEs, etc.), infrastructure management practices, and administrator vision. Subjects are evaluated according to such factors as authentication level, access level, abnormality of behaviour, etc. Objects are represented by such factors as the level of vulnerability, the frequency of access, the level of data sensitivity, etc. Interaction factors such as attack vector or network type are also suggested. These are all combined, to determine risk levels considering probability of an attack and the level of seriousness of the attack.

**Tool Name:** Access Control Risk Evaluation Using Fuzzy Logic. **Purpose:** To evaluate access control adequacy using fuzzy logic. **Features and Benefits:** The mathematical model of the methodology is based on fuzzy logic, which allows to consideration of ambiguity in the description of the state of the system and assessments of its vulnerabilities. **Uniqueness:** By combining modern network monitoring tools, modern vulnerability libraries and comprehensive IT infrastructure data analysis using fuzzy logic, we achieve a more objective and effective risk assessment. This combination outperforms other existing approaches and methodologies or analyses that lack any of these components (Lytvyn et al, 2024). **Input:** Influencing factors, indicators or events observed from logs or other TELEMETRY tools. An example of one indicator is the determination of the frequency of access to objects via monitoring logs from servers and network equipment, setting up connection logging or port activity monitoring, which may be collected, filtered and analysed using the SIEA system (described later) to determine absolute and relative values of the frequency of access to the service and determine the term of the appropriate linguistic variable for the level of the factor. **Output:** In the system of rules forming the knowledge of the system, the influence of this indicator is combined with other factors, such as the criticality and vulnerability of objects, for a comprehensive risk assessment. For example, if an object has a high rate of access and at the same time is critical to the system or has known vulnerabilities, the risk for this object is considered increased. **Current status:** Initial experiments are complete, as described in Lytvyn et al. (2024). The results of testing the proposed methodology can help to improve the applied access policies and the access control system itself, including at the human-machine level. **Lessons Learned:** This approach allows to estimate how often an object is used, which can indicate its importance or potential vulnerability, but how exactly it will affect and what its meaning says will be clarified further in combination with other factors in the mathematical model. Further development of the work includes expanding the pool of indicators to increase the application potential of the methodology, integration with widely used network management tools, and increasing the adaptability of the methodology through testing on real IT structures. Also considered is the improvement of the mathematical apparatus for more accurate estimates and the creation of a base for applied AI tools.

# Runtime Monitoring, Analysis & Detection Tools

## Nokia Anomaly Detection Pipeline

The 'Nokia Anomaly Detection Pipeline' (NAD) aims to ingest sensor readings from IoT devices to predict anomalous behaviour of these devices in near real-time. It is based on training a machine learning (ML) model during normal operation of the DUT that describes expected sensor readings from the DUT: i.e. a fingerprint of the device. When this ML model is then applied during operation, any deviation from the expected sensor readings can be detected and reported.

**Tool Name:** Nokia Anomaly Detection Pipeline. **Purpose:** The Nokia Anomaly Detection Pipeline (NAD) uses past sensor measurements of a DUT to make predictions of expected behaviour during live operation, and compares the prediction (target) with the measurement (actual) and makes a decision to report significant deviations in near-real-time to downstream components. **Features and Benefits:** This tool takes a fingerprint of recorded sensor readings from the DUT operated during normal operation and can detect abnormal situations during live operation. **Uniqueness:** The Nokia Anomaly Detection Pipeline contains a machine learning application, i.e.: the AMB, which is tuned to predict the values of a chosen key performance indicator (cKPI) based on values of other key performance indicators (KPIs). In this way, the tool is capable of detecting deviations not only from the cKPI but also from the other KPIs which are used to predict the cKPI. **Input:** Numeric values of measurements from the DUT reported in regular intervals. **Output:** Alarms containing information on when an anomaly is detected mapped to an alarm condition scale (off -> yellow -> orange -> red) which indicates how long the anomaly persists. As soon as the DUT operates normally, the alarm condition is de-escalated (red -> orange -> yellow -> off). The output information also contains timestamps when transitions in alarm conditions occur. **Current Status:** A feasibility study of the effectiveness of the NAD pipeline has been successfully launched in the Smart Manufacturing use case (UC2, covered later), when monitoring the sensor readings of a robot in operation. The tool chain has been tested in the use case for both stages of operation: the training phase and the (model) application phase. First results of the NAD Pipeline when the robot is forced to operate in anomalous speeds look promising. We can detect anomalies in the reported speed of joint 1; any deviation in the 6 reported angles of the robot; and anomalies in a fraction of the operation cycle of the robot. **Lessons Learned:** It is a known fact that the quality of the training data impacts the performance of an ML-model significantly. We identified irregular gaps in the robot data when analysing plots and tables of this data. Normally, such gaps would be reported as anomalies. To overcome this problem, we crunch the robot data over a pre-configured timeslot via timeseries processing. With a timeslot of 2 or 3 seconds, the resulting data showed no gaps anymore and can be used for model building and model application.

## Misuse Detection ML Toolkit

The *Misuse Detection ML Toolkit* is a set of runtime libraries that includes several AI/ML algorithms with the capability of training ML models. These ML models will be trained to detect the misuse by humans of software components & systems, developing ML for Intrusion Detection (Lee & Stolfo, 1998, Vinayakumar et al, 2019, Stibor et al. 2005) towards misuse detection based on baseline normal behavioural patterns. These patterns would be identified in historic usage scenarios such as normal activities of users and/or similar patterns on log files or data storages. The ML will learn from user-interaction and the detection of divergences in user behaviour from the norm, using similar principles to social engineering for capturing user aspects such as user functional footprint, temporal behaviour and statistical data distribution. The Toolkit is composed of three main modules: *Model trainer*: allows training of AI/ML models; *Model manager*: allows the management and deployment of trained models; *Execution runtime*: allows the execution and serving of deployed trained models.

**Tool Name:** Misuse Detection ML Toolkit. **Purpose:** To train ML algorithms and execute the trained ML models so that the misuse of software components & systems can be detected. **Features and Benefits:** The Toolkit allows users with little or no analytical knowledge to train their own AI/ML models and deploy them in a central platform ready to be incorporated into either bigger and wider applications as a submodule or as a standalone callable service. **Uniqueness:** The Toolkit allows users with little or no analytical knowledge to train their own models by following a wizard approach that will guide the user across screens and options depending on the algorithm selected. **Input:** The input that the *Model trainer* is prepared to receive is any type of sensor-related (numeric-based) data from the DUT / SUT. The *Model manager* is capable of managing and deploying trained AI/ML models regardless these have been trained with the *Model trainer* or not. When executing a trained AI model through the *Execution runtime*, the input that any AI/ML trained model can receive will be sensor-related data from the DUT / SUT. These data will be relayed by the *Execution runtime* to the pod where the trained model will be containerized. **Output:** As output, the *Model trainer* provides the trained AI model, that can then be deployed in the target system through the *Model manager*. *Execution runtime* publishes alerts referring to the anomalies detected. **Current status:** The first version of the Model Trainer is developed and is being tested by the Smart Manufacturing use case (UC2) where information from sensors is gathered and compiled all in a JSON file for triggering the training phase of the use case. For the first attempts to train the UC2 datasets, a two-layered Convolutional Neural Network (CNN) from Keras is being used with 64 filters, activation *relu* and a hidden layer based on MaxPooling1D and GlobalMaxPooling1D configuration so that the normal behaviour of the robot can be modelled, and the anomalies fabricated by UC2 can be used to validate the training of the dataset. The next development to be included here will be the ability to read SHAP-based charts that explain how the model works so that the final user is able to understand the model trained. **Lessons Learned:** The model that was initially trained for UC2 has evolved by doubling the initial layers and adding more filters. In addition, continuous initial data collection has led i4RI and its Data Scientists to adapt the model to the changes in structure and frequency of the data provided.

## r-Monitoring - Monitoring & Analysis of System Processes, Metrics and Network Traffic

The tool aims to enhance system security by providing comprehensive monitoring and analysis of system processes, metrics, and network traffic following patterns suggested by Shao et al. (2010). It includes dynamic file monitoring capabilities, which track changes to critical system files and directories in real-time. Any unauthorized or suspicious modifications are flagged and alerted to the system administrator as these could be indicative of malicious activities. Additionally, the tool continuously scans and evaluates running processes against known malware signatures and anomalous behaviour patterns to identify potential threats, ensuring proactive threat detection and response. This multifaceted approach fortifies the system's resilience against a wide array of security threats.

**Tool Name:** r-Monitoring Tool. **Purpose:** Comprehensive monitoring and analysis of system processes, monitoring metrics, file monitoring. Signature checking and network traffic. **Features and Benefits:** *Resource Monitoring* (CPU & memory consumption); *System Monitoring* (detailed process consumption); *File Monitoring* (monitor sensitive files for changes); *Hash Monitoring* (check hashes for altered processes); *Network capturing and Monitoring* (signature monitor); *Anomaly Detection on Network Data*; Record *Historical Data*; output to Dashboard. **Uniqueness:** The agent designed for collecting metrics, system parameters, and network traffic boasts a lightweight architecture that is compatible with various CPU architectures. Its efficient design makes it especially well-suited for devices with constrained resources (e.g. IoT), ensuring broad applicability without compromising performance. While several tools on the market provide some of the features listed above (e.g. Casola et al, 2019, Ghanem et al, 2013), it is rare to find a single tool that encompasses these capabilities comprehensively and even more for devices with constrained resources. **Input:** configuration in terms of system metrics, file status, network traffic to be monitored by the agent. **Output:** system metrics, assessments of unauthorized or malicious modifications & activities, and alerts for abnormal behaviour of system critical parameters. **Current Status:** The initial version of the monitoring agent has been successfully developed and tested on IoT devices, specifically ZTE router provided by the Telecoms Use Case (UC3). This agent effectively captures key metrics and network traffic data, which it then forwards to the monitoring application for further analysis. **Lessons Learnt:** IoT devices are often constrained by limited resources, necessitating the development of the monitoring agent in a low-level language to efficiently manage these restrictions. Additionally, the diverse architectures present in IoT devices require thorough exploration and investigation. This is crucial for developing techniques that enable the application to operate seamlessly across all devices, achieving broad compatibility and performance optimization in a variety of hardware environments.

## r-Anomaly Detection

This tool is designed to monitor network traffic and identify unusual patterns that deviate from established norms, using intrusion detection approaches exemplified by Sommer and Paxson (2010) and Fuentes-García et al. (2021). It utilizes a sophisticated algorithm to analyse a dataset representing healthy or typical network activity. By continuously comparing incoming traffic data against this baseline, the tool efficiently flags anomalies, which could indicate potential security threats or system failures.

**Tool Name:** r-Anomaly Detection. **Purpose:** This tool is designed to monitor network traffic and detect anomalies by analysing incoming data against a predefined baseline of typical activity. It employs machine learning algorithms (see e.g. Wang et al, 2021) to identify deviations and the underlying causes of these irregularities, that may suggest security threats or system malfunctions. The purpose is to enhance network security and reliability by promptly flagging potential issues. **Features and Benefits:** Identify patterns that deviate from the norm (As established by a provided dataset of typical activity). Pinpointing the specific features that contribute to each detected anomaly (e.g. Lundberg & Lee, 2017). **Uniqueness:** The r-Anomaly Detection tool is dedicated in interpreting network traffic, then detecting and assessing protocol-specific deviations (regarding the protocols used and monitored) from the reference datasets. **Input:** Network traffic, which consists of data packets, each containing parameters from various layers. **Output:** JSON object that contains three values: one indicating whether an anomaly was detected, one describing the severity of the deviation from the defined baseline and a third defining the list of parameters that are considered abnormal. **Current Status:** The current status is that we are analysing the data provided by Smart Manufacturing (UC2) and Telecommunications (UC3) to extract valuable insights. These insights are essential for designing the architecture of our model, and ensuring it is tailored to meet specific requirements and performance. **Lessons learnt:** Network data is characterized by a variety of protocols, each with distinct features. Exploring and investigating methods to handle this diversity and anomalies occurring in each case presents a significant challenge. Tackling this challenge has taught us the importance of flexibility in model design to adapt to varied data types. Our pursuit to develop a robust model has emphasized the need to investigate advanced analytics and machine learning techniques. These methods are crucial for effectively handling and interpreting diverse network traffic, and for assessing its divergence from baselines, with the aim of improving reliability and consistency in our outcomes.

# Risk Modelling of System / Device Under Test

## Spyderisk System Modeller - Risk Assessment

The Spyderisk System Modeller, abbreviated to SSM (Phillips, 2024) is a comprehensive automated risk management toolkit designed to enhance a system's security via the assessment of risks and recommendations of controls to lower the likelihood of risks with an unacceptably high level. SSM enables users to construct detailed system models (models of the SUT / DUT), identify cybersecurity and compliance risks, and determine the most effective mitigations. Its user interface is shown below in Figure 4.

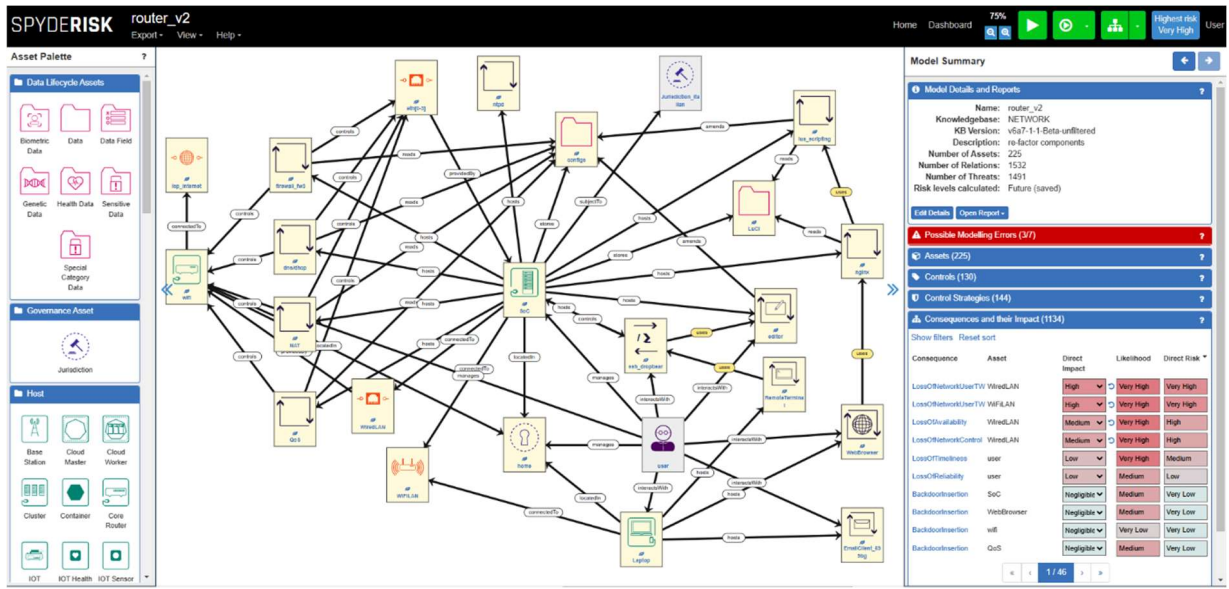

Figure 4: Risk Modelling User Interface

A panel of asset icons is shown at the left, from which a system model can be constructed in the central panel, and at the lower right individual risks (named Consequences) are shown. Each risk can be explored to determine threats that cause it, and controls are recommended by the tool. The controls can be selected and the risk levels re-calculated to see the effect of the controls on the risks. The SSM also supports dynamic cybersecurity risk assessments that adapt to new threats as they are detected by tools like vulnerability scanners (e.g., Wazuh[10]), plus other tools developed in TELEMETRY such as anomaly detection, fuzzing or the event pipeline. This adaptive feature enables the assessment of emerging risks at runtime along with by dynamically recommending and implementing necessary security controls, so they be addressed in a timely manner, enabling the maintenance of a robust defence.

**Tool Name:** Spyderisk System Modeller (SSM). **Purpose:** A comprehensive automated risk management toolkit designed to enhance a system's security via the assessment of risks and recommendations of controls to lower the likelihood of risks with an unacceptably high level. **Features and Benefits:** It is a knowledge-based risk assessment tool enabling system-level risk assessment at both design time and runtime. SSM system model concepts are compliant with the ISO standards 27000/27005. SSM enables users to construct detailed system models (models of the system under test), identify cybersecurity and compliance risks, and determine the most effective mitigations. **Uniqueness:** Spyderisk enables modelling and analysing risk in information systems, based on ISO 27005, part of the ISO 27000 family for information security. It is at TRL6 and has been in development for 9+ years. Knowledge of vulnerabilities, threats, risks and controls to manage risk are encoded in a knowledge base designed to support automation using a cause-and-effect approach to risk modelling. The user constructs a model of their system under test, and the tool uses its knowledge base to identify relevant risks and threats and calculate risk levels, and to recommend controls to lower the likelihood of risks. It differs from existing methods of risk assessment in that other methods are largely based on checklists and human judgement, whereas Spyderisk is a simulator of the causes and effects of cyber threats in interconnected systems. Dynamic context and threat propagation via interdependencies of assets and consequences offer a continuous monitoring and updating of risk assessments as new threats emerge or as changes occur in the system. **Input:** model of SUT / DUT, state reports about system vulnerabilities, typically including CVE (Common Vulnerabilities and Exposures) and CVSS (Common Vulnerability Scoring System) metrics. **Output:** a JSON object that presents the current status of the risk model and includes a list of potential recommendations. Each recommendation specifies the controls that need to be implemented and details the expected risk reduction resulting from the implementation of these controls. **Current Status:** SSM is deployed in Smart Manufacturing (UC2) and a process of mapping the indicators from tools such as the anomaly detection tools to vulnerabilities in components in the risk model is underway. The SSM is also deployed in Telecommunications (UC3), where a risk model of the DUT (a domestic modem-router, known as the Residential Gateway) has been created. **Lessons Learned:** it is necessary to consider modelling devices as systems, and a model of the router DUT has been created as a system in its own right, comprising constituent components of the router such as NAT, routing tables, switching hardware, DNS / DHCP, firewall, wifi module, ethernet connection, http server for configuration, etc. This work has been informed by the SBOM generation tool that has analysed the router DUT and provided an SBOM. The SBOM provides information of libraries and components with versions, which can then be linked to CVEs databases, and therefore vulnerabilities in the databases can be linked to the SBOM. These vulnerabilities are needed to be reflected in the risk model, so a key lesson learned here is that there is a need for mapping of the functional blocks in the risk model to the libraries / components from the SBOM and to understand the parameters in the risk model that are undermined should a CVE be detected. The SSM uses CVSS (currently v2 is supported but extensions to v4 are underway) to represent vulnerabilities in the risk model, which facilitates this mapping, but the mapping is highly domain and application specific.

---

[10] https://wazuh.com/

# Testing / Evaluation Environments

## Digital Twin

A Digital Twin acts as an infrastructure emulation tool designed for conducting varied security tests on IoT devices in a customised and isolated environment, with a primary focus on assessing the security of their software components. It utilizes a firmware file as input to emulate the capabilities of the IoT device, with a specific emphasis on system and configuration files. Its main objective is to assist in identifying vulnerabilities related to firmware, IoT device configuration, and system OS components.

**Tool Name:** Digital Twin. **Purpose:** Provides an emulation environment, allowing IoT developers to create an isolated testing platform and test the software stack including the operating system and device drivers. **Features and Benefits:** The tool offers benefits as a virtualized infrastructure, enabling other security tools and testing (for example, SBOM) to operate without the need for physical hardware. **Uniqueness:** The tool is constructed using the open-source quick emulator (QEMU) package, which offers a versatile framework for emulating diverse hardware platforms. this framework is enhanced by facilitating the tool's ability to effortlessly replicate IoT device configuration files and Integrating vulnerability scanners to detect & analyse security vulnerabilities effectively. **Input:** Firmware files for IoT devices, which can be obtained by downloading it from the IoT vendor's servers or by extracting it from the device using reverse engineering methods. **Output:** N/A. **Current Status:** Firmware imported to QEMU is analysed using the "binwalk" tool, and the extracted files are manually studied to determine the root file system. Printable strings are extracted to help identify libraries used (see also SBOM tool). **Lessons Learned:** This part of the toolset is still in an early stage, but it is clear that there is a wide variety of firmware and each twin will need customisation to the unique challenges of the target firmware.

## Cyber Range

TELEMETRY uses a Cyber Range in the form of a commercial platform developed by AIRBUS[11] for the development, delivery and use of interactive cybersecurity hybrid environments that can include both virtual and physical components. This platform is available to project partner Munster Technological University (MTU) and can be used for testing, scenario development, data extraction, attack simulation, etc. It can be used to generate simulated network attacks as test scenarios for the anomaly detection tools.

**Tool Name:** Cyber Range. **Purpose:** Cyber Range is a platform for the development, delivery and use of interactive cybersecurity hybrid environments. **Features and Benefits:** The Cyber Range is a platform with a variety of features around cyber security testing. it can generate a variety of simulated traffic and attack patterns that can be applied to an SUT. **Uniqueness:** The Cyber Range supports a wide range of cyber security testing features. Even though it will not be physically moved to one of the Use Case testbeds, it can connect to them remotely via suitable methods such as VPN and can perform tests through that remote connection. **Input:** A description/specification of the SUT is needed to create traffic patterns that simulate potential attacks. **Output:** The Cyber Range exposes the SUT to simulated attack traffic in order to be able to investigate the SUT's response and the detection by anomaly detection tools. **Current Status:** The Cyber Range is available on MTU's premises. Test scenarios and attack patterns will be defined for the TELEMETRY use cases. **Lessons Learned:** A twofold role has been identified for the Cyber Range in the context of the project: Firstly, the use as a traffic generator to emulate attacks to the SUTs. This requires a VPN link to the use case testbed. In addition, the Cyber Range will be used to run tests on elements of the DLT based data space that is part of the TELEMETRY architecture. This is intended to ensure that the TELEMETRY infrastructure itself does not create vulnerabilities and does not open new attack vectors on the SUT.

---

[11] CyberRange: Advanced simulation and training solution, https://www.cyber.airbus.com/products/cyberrange/

## Dedicated Testbeds

Dedicated test environments simulating real-world conditions have been built as part of the TELEMETRY framework. An example of this is for UC3 in Telecommunications. This included connecting physical Residential Gateways (RGWs, also known as modem-routers) to both an internal local network (simulating a home environment) and the external Internet. We also added a specialized traffic generator to the internal network, whose role is to reproduce the typical behaviour of users and the applications they use, in order to generate a baseline of legitimate and realistic traffic, needed for training machine learning based anomaly detection tools. The generator is also able to inject attacks or simulate the control phases of compromised hosts. This includes sending packets with spoofed source IP addresses (e.g., for reflection and amplification attacks), executing exploits against vulnerable hosts, establishing cover channels to exfiltrate data (e.g., DNS and ICMP tunnels.), creating sessions of anomalous duration, generating traffic during unusual time slots, or, in general, communicating with unknown IP addresses and domains. TELEMETRY tools are also deployed on the LAN side to collect and monitor traffic data. Additionally, they can capture WAN traffic from the external interface of the RGWs when necessary. In general, the testbed was designed and set up with the primary aims of:

- Reproducing a realistic network environment with typical traffic patterns.
- Hosting the TELEMETRY tools and sensors for development and optimization.
- Providing an environment for effectively evaluating TELEMETRY tools' ability to test RGWs and generate accurate results.

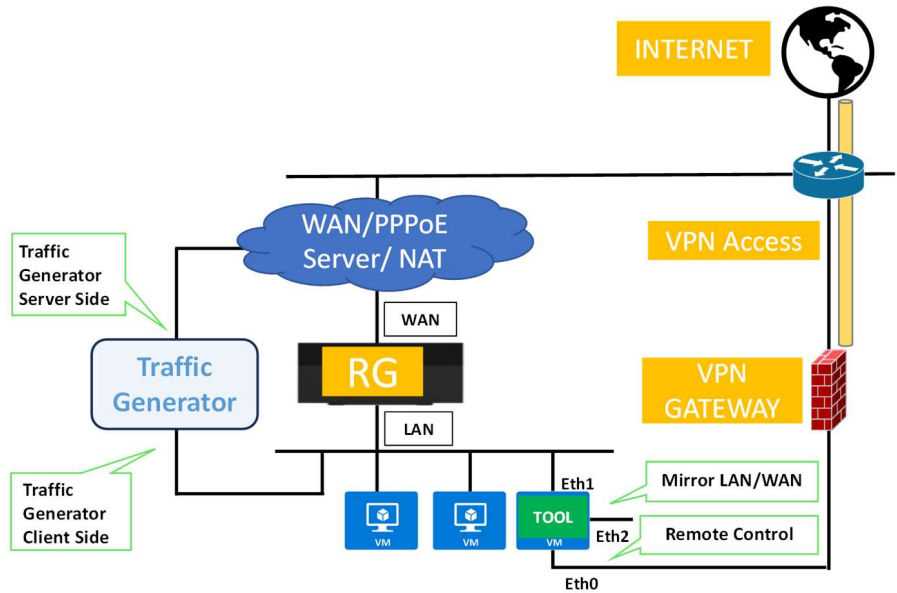

Figure 5: Telco Use Case Testbed

# Security Controls

## Secure Software Updates

This tool perform efficient and secure software updates of the SUT / DUT, utilizing lightweight cryptographic algorithms and novel protocol solutions to develop a new framework for secure patches.

**Tool Name:** Secure Software Updates. **Purpose:** The Secure Software Updates tool aims to provide a robust and efficient mechanism for performing secure firmware and software updates on distributed and resource-constrained devices, such as IoT devices. Given the diverse range of communication protocols and the challenges associated with lightweight devices, our approach focuses on minimising computational overhead while ensuring the highest level of security. This tool will be crucial in maintaining the integrity and functionality of systems in decentralised environments, where traditional methods fall short due to the limitations of low-power and low-bandwidth networks. **Features and Benefits:** The Secure Software Updates tool is more than just a utility; it is an enabler for the entire TELEMETRY architecture. By integrating novel lightweight cryptographic algorithms with optimised protocol solutions, the tool ensures that software patches and updates can be distributed and applied quickly and securely, even in the most resource-constrained environments. The primary benefits include: (1) Lightweight Operation: The tool leverages efficient cryptographic MAC algorithms and streamlined key management, which significantly reduces the energy and computational costs associated with secure updates. (2) Seamless Integration: The tool is designed to be easily integrable with existing IoT systems. It supports a wide range of communication protocols and can be implemented as a simple firmware update, minimising disruption to ongoing operations. (4) Scalability: With its focus on distributed and decentralised systems, the tool is capable of handling the complexities of managing secure updates across a large number of devices, each potentially requiring different access controls and update protocols. **Uniqueness:** The uniqueness of the Secure Software Updates tool lies in its ability to balance security and efficiency in environments where resources are extremely limited. Unlike existing solutions, which may focus on either security (Duttagupta et al, 2022) or lightweight operation (Duttagupta et al, 2023), this tool achieves both. It accomplishes this by employing innovative cryptographic protocols specifically designed for low-power and low-bandwidth scenarios. Additionally, it addresses the challenge of dynamic access control, which is often overlooked in traditional update mechanisms, by allowing for adaptable permissions that can evolve as the network grows or changes. **Input:** The tool receives inputs from various components within the TELEMETRY architecture, including device status reports, update packages, and security parameters from the PKI. These inputs are processed to ensure that only validated and authenticated updates are distributed across the network. **Output:** The primary output of the Secure Software Updates tool is a secure, encrypted update package that can be deployed across the network with minimal disruption. This tool will be deployed in the form of a cryptographic library, which can be easily called during run-time. **Current Status:** The current status of the project involves the ongoing development of the cryptographic library that will underpin the Secure Software Updates tool. Initial prototypes have been developed, focusing on optimising the balance between security and computational efficiency. We are also conducting a series of tests using simulated IoT environments to refine the tool's performance under various network conditions and device constraints. **Lessons Learned:** Developing a secure update mechanism for resource-constrained devices has highlighted several challenges (Alrawi et al. 2019). One key lesson is the importance of adaptability; our approach must be flexible enough to handle diverse network conditions and device capabilities without compromising security. Additionally, managing the trade-off between security and efficiency remains a constant challenge, particularly in ensuring that the cryptographic algorithms used are both lightweight and robust.

# Framework Infrastructure

## Data Space - DLT based Data Sharing

As shown in the TELEMETRY architecture above, a data space is envisaged as an important element for data sharing among the tools, and as a place where a record of events, test results and actions is maintained. With the variety of tools contributing to the overall TELEMETRY solution, the data space serves as a common repository where records are being kept of what was reported by the tools. In order to ensure that the records are reliable, trustworthy and meet requirements for auditability, they have to be resilient against manipulation. To address this TELEMETRY proposes a Distributed Ledger Technology (DLT) based data space as a data sharing and persistent storage solution for its tools ecosystem. The immutability feature of DLT facilitates tamper-proof records of events and actions. Its distributed nature adds resilience, and it also allows tools in different locations to contribute to the same records and share data such as test results or alerts among themselves.

The DLT based data space in TELEMETRY extends prior work on a platform known as SmartQC (McGibney et al, 2024) which was developed in a smart manufacturing context. SmartQC provides a JSON API for access to the data space and facilitates the definition of context and meta data structures that can be extended and validated prior to being committed to an underlying DLT layer. Any data that is committed to the ledger has to adhere to the defined contexts and metadata structures. In the scope of TELEMETRY, a data space based on an extension of SmartQC is used to create immutable, auditable records of events, actions, indicators and reports that are generated by other tools.

**Tool Name:** Data Space. **Purpose:** Data sharing platform based on DLT and Smart Contracts, for maintaining an immutable, auditable record of events, alerts and reports generated by the various security and anomaly detection tools. **Features and Benefits:** The platform offers the benefits of DLT, in particular immutability of records shared in a distributed ledger, while abstracting from its complexity by featuring a data sharing API on top that is accessible by the various tools in the TELEMETRY ecosystem. **Uniqueness:** The tool provides a JSON API on top of an underlying DLT platform, offering an abstraction layer that facilitates the interaction with the data sharing platform through relatively simple JSON messages while making use of DLT features such as immutability in order to create reliable, auditable records. The abstraction layer further allows for flexibility in the choice of underlying DLT based on the features that are required, instead of being limited to one particular platform. **Input:** In relation to testing tools such as the Network Fuzzer or the SBOM Generator, events and reports will be received directly from those tools. In the live operation stage, aggregated reports originating from anomaly detection tools will be received via the SIEA pipeline. **Output:** Through the Client API, tools such as the SSM or the Trust and Security Analyser can query the data sharing platform for reported data. Notifications will be triggered by incoming reports from testing phase tools and will be sent to inform the SSM of new content.

**Current status:** The existing platform SmartQC, which serves as a basis for the data space, was developed in an earlier project. It can accommodate a selection of underlying DLT flavours such as Hyperledger Fabric (Androulaki et al, 2018), IOTA, and BigchainDB (McConaghy et al, 2016). This SmartQC platform is modified and extended to address the requirements of TELEMETRY. **Lessons Learned:** In relation to functionality, one particular requirement that demands an extension to the current platform is the need for actions or notifications that are triggered by certain transactions. When new reports are received from testing tools such as the Network Fuzzer or the SBOM Generator, these may be of relevance to the risk assessment that is being done by the SSM. Hence, the SSM has to be notified of those new reports, meaning that the transaction that commits the report to the ledger should at the same time trigger such notification. In addition to the extension of functionality, underlying context and meta data structures will be defined in line with the content and semantics of output generated by the TELEMETRY tools, and also aligned to expected inputs towards the risk analysis. This is done in alignment with cybersecurity indicator specifications that are developed in the project, so that there will be a mapping between indicator specifications and the data structures that are to be used in the data space.

## Security Information and Event Acquisition (SIEA) Pipeline

In the TELEMETRY framework, multiple upstream anomaly detection tools will report their findings during operation. It is a task of the 'Security Information and Event Acquisition' (SIEA) pipeline to ingest, aggregate, filter and prioritize these security related events and forward compact results to the data space or downstream tools. This is necessary because some events (e.g. repeated status messages with the same information but only separated by time) flood the message exchange, effectively creating a benign or inadvertent denial of service attack. Therefore, there is a need to filter, aggregate or summarise these events.

The SIEA pipeline (Figure 6) consists of a set of applications: The *Aggregator* subscribes to a message broker and listens to alerts and insights reported by the upstream tools. The Aggregator is aware of the priorities mapped to each security-related event and can therefore control the flow of data to the Distributor. Next to this feature, the Aggregator can enforce a 'silence period' if multiple alerts of the same type are reported in high frequency by a 'noisy' upstream tool. The *Distributor* is a non-blocking web service capable of buffering events from the Aggregator in case a downstream tool is not ready for ingesting the next chunk of reports. The *SIEA* itself engages in the handshakes of downstream tools guaranteeing the correct processing of the compact reports without overloading the tools themselves. The applications are supported by a database which stores the status of the downstream tools and the corresponding SIEA application for forensics.

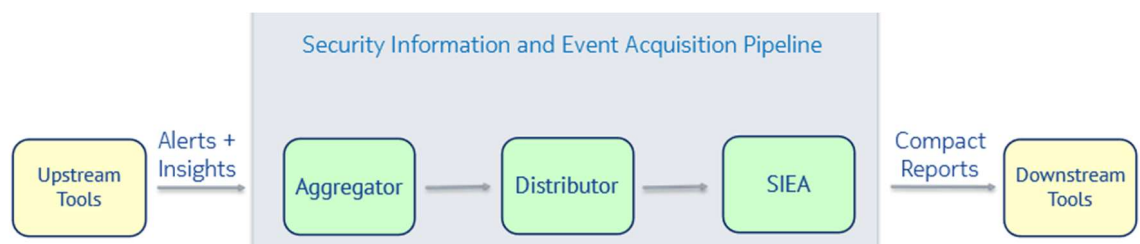

**Figure 6: SIEA Pipeline**

**Tool Name:** Security Information & Event Acquisition Pipeline. **Purpose:** Collects, aggregates, filters, and prioritizes security-related events and vulnerabilities. Connects to the downstream components via RESTful southbound interface for further processing. **Features and Benefits:** The SIEA pipeline subscribes to a Kafka broker which publishes events from upstream tools without changing their order. These events are pushed into three queues of different priorities, i.e. high, medium and low. Less prior events can be 'overtaken' by higher prior events. Example: Low-prior events are only forwarded downstream, if the high and medium queues are empty. In order not to flood the downstream tools, the SIEA pipeline groups, aggregates and, in some cases, suppresses recurring events. The SIEA pipeline controls the flow of events to downstream tools via RESTful handshake. **Uniqueness:** The SIEA Pipeline operates in real-time and offers other upstream tools a controlled connection to its consumers. **Input:** Security related events (from e.g.: Wazuh Server, Snort, etc). The Aggregator listens to a set of Kafka topics, each of which is dedicated to an upstream tool. The semantic descriptions of the JSON messages which are injected by the upstream tools must be provided by the tools themselves. **Output:**

Prioritized, aggregated and filtered events specific to the upstream tools - the pipeline only adds additional information to the JSON e.g. its priority, a potential repetition count and additional fields which the SSM might require. **Current Status:** The SIEA Pipeline has been implemented with 4 applications. At this point in writing, all applications are dockerized. The next step is to connect the SIEA pipeline to the NAD pipeline using a Kafka Broker. **Lessons Learned:** The Aggregator must be able to ingest messages from different upstream tools. Based on what they deliver, adaptations are required in the Aggregator itself. Analog to this, the SIEA must perform handshakes with different downstream tools. Since the SIEA is dockerized, the plan is to setup different customized instances of the SIEA for each destination tool. The advantages of this approach are a performance boost and an avoidance of blocking situations.

### Workflow Orchestrator

**Tool Name:** Workflow Orchestrator. **Purpose:** Infrastructure component that configures and executes tools in sequence following workflow templates or user control. **Features and Benefits:** Provides a flexible means of automated tool execution and chaining, enabling different workflows to be specified and triggered either manually or upon specified event conditions. **Uniqueness:** Flexibility of workflow templates, specification of trigger conditions, specificity to the application domain of cybersecurity testing and monitoring, integration with immutable data space. **Input:** workflow template, tool configurations, trigger conditions, control signals from dashboard. **Output:** Automated execution of tools and storage of results & indicators in Data Space. **Current Status:** In development. **Lessons Learned:** The need for workflows has been established as part of the TELEMETRY work on methodologies, which originated as human-executed methods but the types of tools and the frequency of execution, plus reactive execution of components in response to events has led to the requirement for automated workflow execution & management.

# Test Cases

## Use Case 1 – Aircraft

Use case 1 focuses on flight cargo monitoring by Antonov Company. The purpose of this is to enable enhanced monitoring of valuable items in the cargo hold and certified transmission of data for live monitoring. The cargo monitoring system monitors the hold visually, whilst recording pressure, humidity, temperature and g-force in three-axes. Sensors are installed with the cargo and the sensors are managed by a special program for monitoring parameters, which is installed on a laptop located in the hold and connected to the local wireless network in the hold. an overview of the scenario, along with the anticipated threats, is shown in Figure 7.

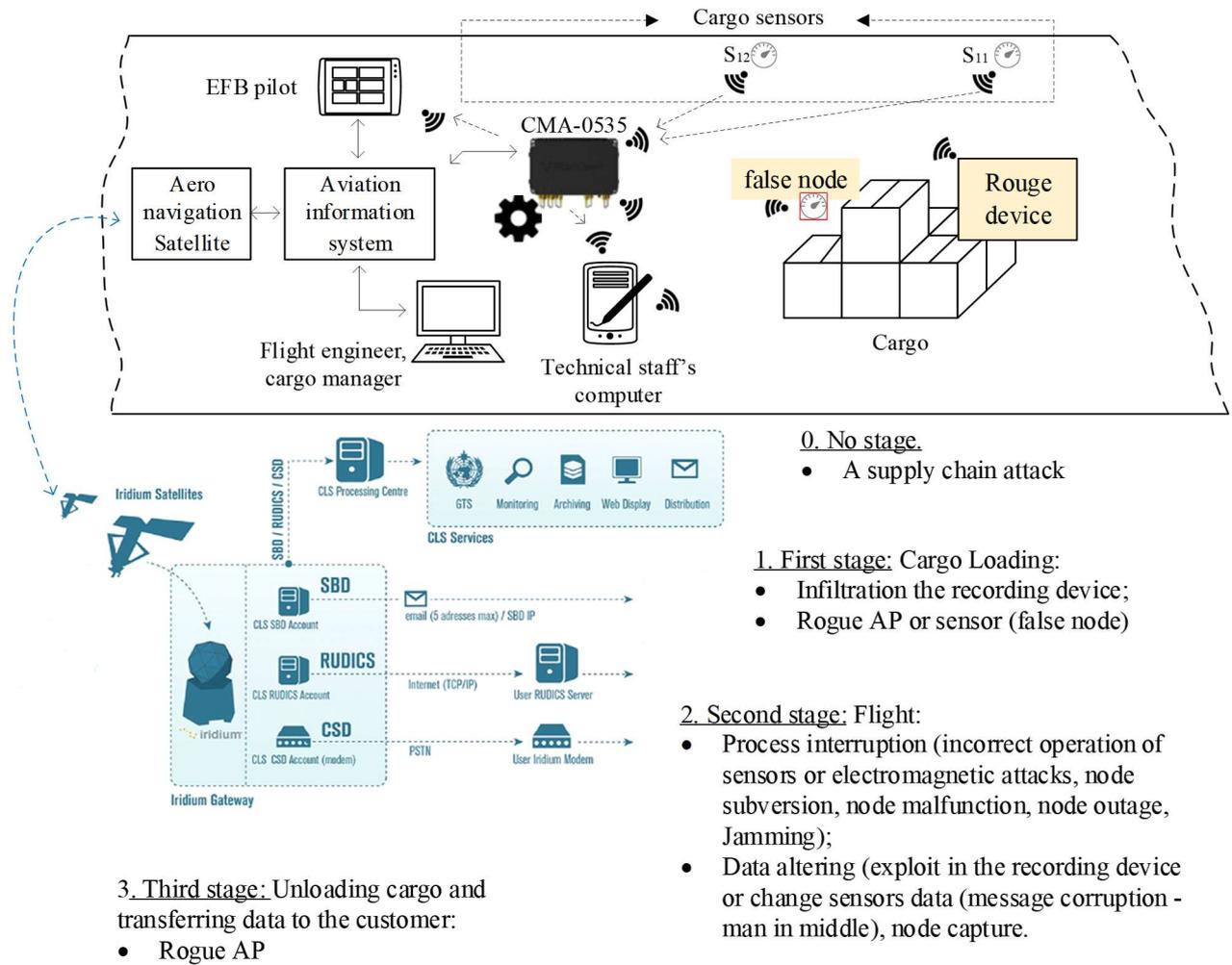

**Figure 7: Overview of Threats in the Aviation Use Case**

## Device Operation Anomaly Detection

TELEMETRY tools can help improve the security of an air cargo monitoring system based on on-board indicators, information traffic analysis tools and anomaly detection. Machine learning tools developed by TELEMETRY partners will help detect atypical behaviour of onboard information sensors (temperature, pressure, humidity), and suspicious network activity and track requests to connect to abnormal network access points. The TELEMETRY tools will be installed on a separate laptop residing in the airplane. This laptop is connected to the sensor network in the airplane. It will host all required tools for the wired and wireless approach, as well as for evaluation purposes record the raw data. This allows to evaluate the tool events with respect to real events. The workflow is shown in Figure 8.

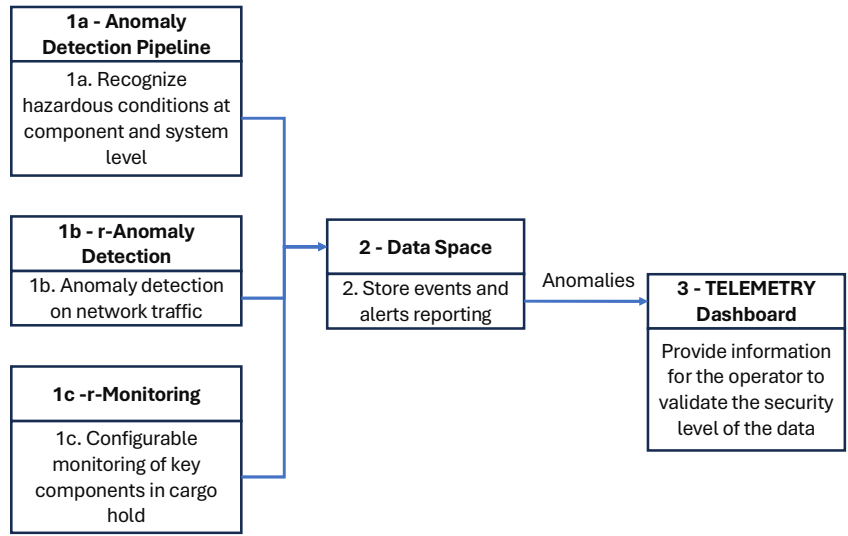

Figure 8: Anomaly Detection Workflow

## Access Control Risk

This scenario will identify vulnerabilities and risks in access control to system components that regulate the creation of temporary or permanent users with different access levels and sets of rights, and also allow to control their typical or atypical behaviour. The multiple accesses to information give the user mechanisms to obtain permissions from multiple policies, leading to an accumulation of effective permissions and collectively presenting a certain level of risk. A universal tool for testing access control systems will allow the network administrator to improve the quality of decisions made and reduce response time to incidents. Its workflow is shown in Figure 9.

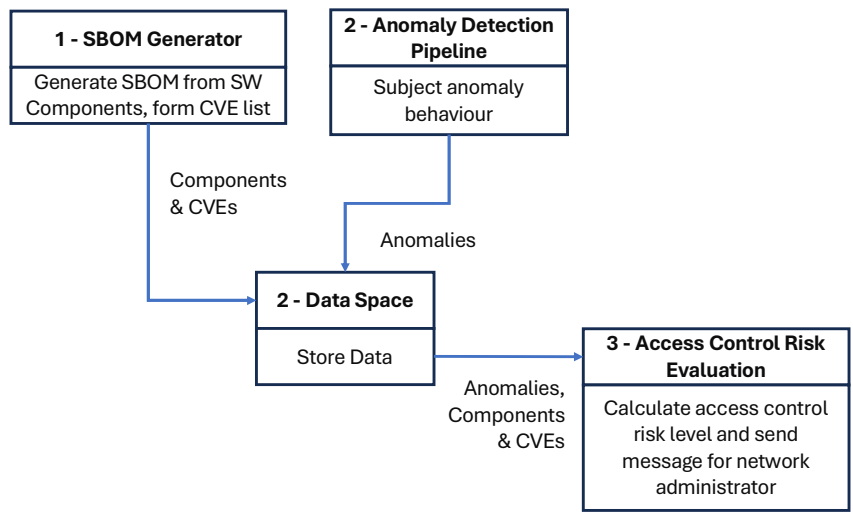

Figure 9: Access Control Risk Workflow

## Use Case 2 – Manufacturing

Smart factories consist of various interconnected components, such as production machines, network devices, IoT devices, and user devices like laptops and computers. These components, being 'smart' and 'online,' are vulnerable to cyber-attacks or could potentially become attackers themselves. Examples of such threats include outdated IoT devices being hijacked, compromised production machines transmitting confidential data, or unauthorized devices infiltrating the network.

Threats include:

- IoT devices with outdated firmware being hijacked.
- Compromised production machines becoming attackers, possibly via supply chain attacks.
- Unauthorized devices entering the network and launching internal attacks.

The approach to addressing these threats is shown in Figure 10.

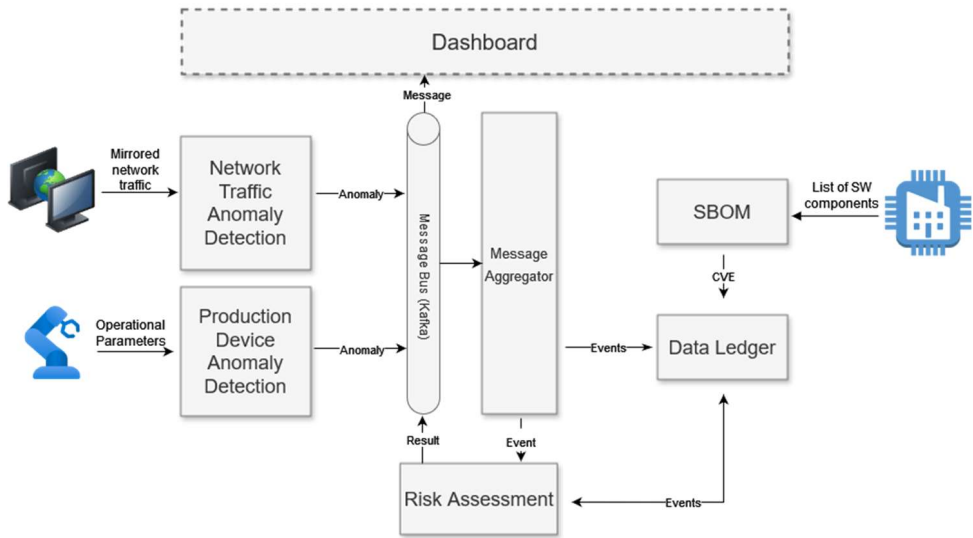

Figure 10: Use Case 2 Architecture Diagram

Current defence methods like firewalls and traffic scanners are mostly static and only effective against known threats, requiring manual updates. To enhance these defences, TELEMETRY aims to use machine learning to detect unknown cyber-attack variants in real-time. For example, machine learning models will analyse robot operating parameters and network traffic to spot anomalies. Additionally, a Software Bill Of Materials (SBOM) will be compared with vulnerability databases to identify potential exploits. A central risk assessment system will compile these findings, offering a comprehensive risk analysis and recommended countermeasures on a dashboard for administrators. To securely document all events that occur within the smart factory, a distributed ledger technology system is also used, ensuring the integrity and traceability of the recorded data. An exemplar workflow for UC2 is shown in Figure 11.

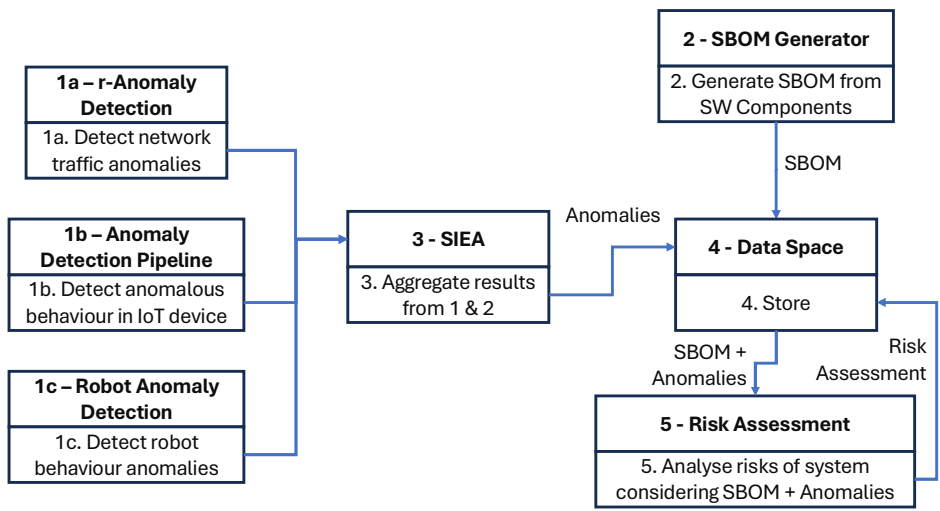

Figure 11: Use Case 2 Exemplar Workflow

## Use Case 3 – Telecommunications

UC3 concerns the TELCO world and in particular the Residential Gateway (RGW) Devices, also known as home routers. These devices play a fundamental role, representing the frontier between the customer domain and that of the operator. The presence of exploitable vulnerabilities in such devices can indeed have serious consequences, in terms of both access to the operator's infrastructure and assets (e.g. VoIP platforms, access control systems, remote management solutions, etc.), and of violation of customer data and systems (traffic interception and redirection, access to internal camera streams, access to file repositories, etc.). Furthermore, with the rise of remote working, these devices handle not only personal data but increasingly often also sensitive professional and business information. It should also be noted that the number of RGW devices distributed and managed by a single operator can in many cases exceed several millions of units. Finally, despite being apparently simple and low-cost, these devices offer a very large attack surface with multiple physical interfaces (fiber, VDSL, Ethernet, WiFi 5GHz, WiFi 2.4GHz, USB, 4G/5G, etc.), active services (DNS, DHCP, SMB, UpnP, DLNA, etc.) and significant numbers of components and software libraries, all of which represent entry points for threats.

In the Use Case under examination, numerous challenges must be addressed to mitigate risks, both from the operator's and the customer's perspective. Two of these stand out due to their significance and representativeness: prioritizing software and firmware vulnerabilities (i.e. CVEs) (vom Dorp etl al, 2022), and mitigating/detecting supply chain attacks (Kotolov, 2021). Since the former refers to known vulnerabilities and the latter to unknown threats, distinct approaches must be employed by the TELEMETRY tools.

## CVE Prioritisation

Through the use of vulnerability scanners and other specialized tools, a significant number of vulnerabilities are commonly identified in RGWs. This is generally ascribed to various factors, such as the huge number of software and hardware components available (applications, services, libraries, drivers, etc.), as well as the often-present outdated components. For example, the best device among the 117 analysed in Weidenbach (2020) was found to be affected by 348 High severity CVEs relating to the kernel alone, whereas the worst by 579 CVEs. Often, fixing all identified issues is unjustifiable with the time & cost constraints imposed by market demands, time-to-market considerations, or the often limited fixing capabilities of the device manufacturer. Consequently, it is essential to have tools and methodologies that can be used to concentrate the available resources on fixing vulnerabilities that maximize risk reduction.

To this end, vulnerabilities are typically classified according to the CVSSv3 methodology, promoted by FIRST[12] and recently updated to version CVSSv4[13]. In the last few years, the CVSS framework, in turn, has been complemented by other approaches which incorporate dynamic parameters into the severity calculation, like the availability of attack code or active exploitation in the wild. An example of those initiatives is the Exploit Prediction Scoring System (EPSS), also promoted by FIRST[14] (Jacobs et al, 2023). Although EPSS is in general a powerful tool for prioritizing vulnerabilities, it presents some limitations and anyway it does not consider the actual deployment scenario, such as the context of the specific Use Case under examination. A first limitation arises from the fact that the RGW firmware provided to the Telco operators, as involved in this Use Case, is typically customized for individual operators and not available as a COTS (Commercial Off-The-Shelf) product. Consequently, identifying network exploits for a specific vulnerability, even if in production, would most likely involve other devices or IT systems with often very different characteristics or configurations, making such a contribution to the severity less reliable. In fact, exploiting a kernel vulnerability on an RGW presents different challenges compared to a traditional server. Moreover, the CVSSv3 and the EPSS primarily focus on individual vulnerabilities, without considering the combined effects that those vulnerabilities can have on the system as a whole. A chain of vulnerabilities in a specific RGW can generate very different consequences depending on the selected starting set.

---

[12] Common Vulnerability Scoring System v3.1: Specification Document, CVSS v3.1 Specification Document https://www.first.org/cvss/v3-1/cvss-v31-specification_r1.pdf
[13] Common Vulnerability Scoring System v4.0: Specification Document, CVSS v4.0 Specification Document https://www.first.org/cvss/v4-0/cvss-v40-specification.pdf
[14] Exploit Prediction Scoring System (EPSS), Exploit Prediction Scoring System (EPSS) https://www.first.org/epss/

In summary, these methodologies, while very interesting and promising for the IT field and for COTS IoT, do not take into account neither the possible relationships or dependencies between the actual device's software modules, nor the access control model (normally in these devices the customers do not have privileged access). For example, a High CVSS scored vulnerability might have a low resolution priority for the Telco operator because it is difficult to exploit it due to the device access control restrictions.

TELEMETRY's approach to addressing this challenge involves primarily the SBOM generation and Risk Assessment tools. The SBOM generation tool provides an SBOM for the router, along with a list of applicable CVEs via query of the database. The Risk Assessment tool has a risk model constructed within it of the router's hardware and software structure, along with external components that reflect a typical operating scenario (e.g. a user's home network with other devices, the telco network and the Internet (Figure 4 above shows a version of this risk model). The CVEs detected are mapped to vulnerability parameters within the risk model and risks are generated. The result is a set of risks with associated risk levels comprised of impact (severity if the risk occurs) and likelihood, plus a mapping from the risk to the threats that caused them, plus the vulnerabilities (represented by CVEs) that facilitated the threats. Therefore, the tool provides a mapping from high level risks to CVEs and allows the CVEs that cause the high level risks, or many risks to be quickly identified. The workflow for this scenario is shown in Figure 12.

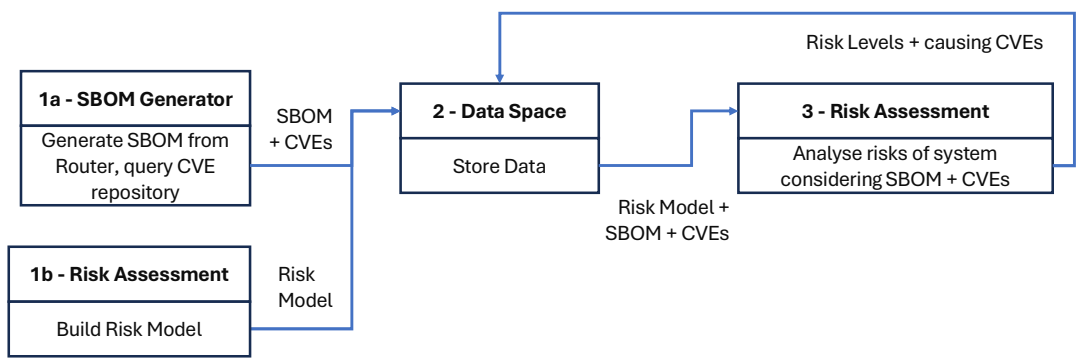

Figure 12: CVE Prioritisation Workflow

## Supply Chain Attacks

A second critical challenge is posed by supply chain attacks. As defined in Kotolov (2021), a supply chain attack is a cyber-attack that seeks to damage an organization by targeting less secure elements in the supply chain. Notable examples include the SolarWinds breach of 2020[15] and the more recent open-source XZ Utils incident of 2024[16], where a fake identity was used to infiltrate the GitHub development team as early as 2021. Device firmware contains a large number of commercial and open-source components. An attack on the manufacturer's infrastructure or a public repository could lead to the inclusion of malicious components on the device which, if undetected, could have serious consequences for the operator and its customers. A malicious component could collect user data, exfiltrate it to external servers, modify device configurations for illicit surveillance (such as altering DNS server IP addresses), compromise internal hosts, and more. Consequently, early intervention during the design/testing phase is crucial, employing tools capable of identifying unforeseen anomalous or malicious behaviours.

This case can be addressed in multiple ways using the TELEMETRY framework, and each has its own workflow.

---

[15] Jake Williams, What You Need to Know About the SolarWinds Supply-Chain Attack, SANS Institute. https://www.sans.org/blog/what-you-need-to-know-about-the-solarwinds-supply-chain-attack/

[16] Cedric Pernet, XZ Utils Supply Chain Attack: A Threat Actor Spent Two Years to Implement a Linux Backdoor, TechRepublic. https://www.techrepublic.com/article/xz-backdoor-linux/

- A first approach can use SBOM generation, where SBOMs determine applicable CVEs for the components within the SUT. A crucial part of the SBOM generation is that it needs to recurse back up the software supply chain – i.e. an SBOM for a DUT is generated, which returns a manifest of components and associated CVEs. Each of these components must have an SBOM generated for its upstream dependencies (i.e. the libraries the component imports), again with associated CVEs. Each of these dependencies must have an SBOM and CVEs found and this process must recurse backwards until no upstream dependencies are found.
- A second approach can use r-Monitoring and Anomaly Detection tools to observe the SUT and warn of any anomalous behaviour in the system.
- These approaches may be combined to provide warnings based on known vulnerabilities introduced by the supply chain, or to detect unknown vulnerabilities and provide warnings.
- Further investigation may be undertaken and correlated with information in SBOMs, results of other tests, risk assessments.

The workflows for each are not illustrated for reasons of space, but a key point that arises from this is there are multiple combinations of tools and that the framework must support this flexibility to be useful to the operator.

# Discussion

## Challenges and Solutions

The challenges posed by Taylor (2024) are addressed by TELEMETRY as follows.

*1) There is a need to consider the* **full lifecycle of IoT components** *– at their design time, their integration into systems, and operation of those systems.* As described above in the relationship between the Microsoft SDL and the TELEMETRY framework, the TELEMETRY tools cover the full lifecycle. Typically testing tools cover design, coding and deployment into systems, and monitoring and anomaly detection tools cover operation of devices within systems, but the tools may be used at any stage of the lifecycle as deemed appropriate.

*2)* **Threats and risks can propagate when components are connected together in systems** *- vulnerabilities in one component can affect other components in a system.* The relationship between the device and the system is addressed via threat propagation in the SSM risk assessment tool, which specialises in transparently describing the effect of devices within systems and impacts vulnerabilities in one component on other components it is connected to in a wider system. The SSM provides analytical tools that enable tracing back from a risk occurring in one component to the threats and vulnerabilities in other components that caused the risk.

*3)* **IoT devices present limitations to current testing and management** *due to geographical distribution, opacity and limited processing power.* TELEMETRY tools address this via lightweight monitoring components that may be installed within low power devices. TELEMETRY also provides multiple tools that monitor systems, the behaviour of devices within those systems and raise alarms if the system behaviour is abnormal.

*4)* **Risk assessment fulfils an important requirement** *because it enables assessment of what elements are important to the system's stakeholders, how these elements may be compromised, and how the compromises may be controlled.* TELEMETRY supports two types of risk assessment: a specialised risk assessment tool for access control and SSM, a cybersecurity risk assessment tool that enables the assessment of devices within systems.

*5) Feedback from operational monitoring of IoT devices can inform firmware updates / patches to the devices but there is a significant challenge in* **rolling out these patches to multiple low-power devices geographically distributed***.* This is directly addressed by the Secure Update Tool, which enables efficient, scalable and lightweight update of distributed and decentralised systems.

## Summary of Key Observations

This work has resulted in further key observations and findings beyond that of Taylor (2024), summarised here.

1. It is a key requirement that the tools can be used in different combinations as decided by the operator. The Framework's architecture is therefore intended to be a flexible, extensible framework aimed at providing testing and monitoring tools for hardware, IoT and software. The framework comprises testing tools; runtime monitoring, analysis and detection tools; risk modelling tools; security control tools; testing and evaluation environments; and framework infrastructure.
2. Tools may be used in multiple different combinations, and ad-hoc use is also encouraged, where one tool may provide clues and other tools executed to undertake further investigations based on initial results. Automated execution of tool chains is supported by workflows representing commonly-used combinations of tools expressed templates describing sequences of tools, user decision points, data interactions, to address a particular problem.
3. For audit and compliance purposes, there is a need for immutable storage of testing configurations, results and other data. This is addressed via storage and sharing using Distributed Ledger Technology (DLT), which facilitates this immutable storage.
4. Indicators are observations or measurements representing information of relevance for assessment of cyber security risk, such as observable vulnerabilities, incidents and threats. An indicator is an output of one tool or a combination of outputs from multiple tools aggregated together (e.g. the correlation of output values may comprise an indicator). Indicators may enable decisions regarding actions or subsequent tooling.
5. Mapping this work to development lifecycles is a useful means of determining applicability and utility of the tools and framework. In this paper we have mapped our work to the Microsoft SDL due to its recency and ubiquity as a methodology from a major vendor.
6. There is a key interplay between devices and systems (DUT / SUT) and the decision whether an entity under examination is a device or a system depends on perspective and context: a device can be a system in its own right and be integrated into a wider system.
7. Anomaly detection in multiple forms is a key means of runtime monitoring. It is implemented using explainable machine learning techniques to pinpoint key indicators. Normal behaviour is identified and recorded, and monitored live observations are compared against it to determine anomaly alarm conditions. A key observation is that deployment each situation requires customisation / training as the normal behaviour is highly specialised to the deployment conditions, and indeed normal is likely to change over time, requiring dynamic model adaptation involving moving norms.
8. Tools' configuration and usage is highly dependent on the SUT / DUT – there is significant complexity and differentiation between e.g. a router's software and hardware construction compared with that of a sensor used within an aircraft, so there will be considerable investigation needed related to the specifics of each device / system as an item of further work.

## Conclusions & Further Work

This paper has presented an overview of the TELEMETRY framework that aims to provide an extensible suite of tools and infrastructure that enables testing and monitoring of ICT ecosystems, with a special focus on IoT devices and their placement within wider systems.

The current status is that the project is approximately one-third through its lifetime. The framework architecture has been specified, use cases are understood and with clear challenges, components are currently in development or evaluation, and approaches to address the challenges in the Background section and the requirements of the project's use cases have been proposed. Further work involves finalisation of tool development and testing each of these approaches, adapting as necessary to the needs and detail of the application area.

# Acknowledgements

This work is within the Horizon Europe TELEMETRY (Trustworthy mEthodologies, open knowLedgE & autoMated tools for sEcurity Testing of IoT software, haRdware & ecosYstems) project, supported by EC funding under grant number 101119747, and UKRI under grant number 10087006.